\pdfoutput=1 
\documentclass[useAMS,usenatbib]{mn2e}

\usepackage[ansinew]{inputenc}
\usepackage{amsfonts}
\usepackage{amsmath}
\usepackage{enumerate}
\usepackage{enumitem}
\usepackage{graphicx,float}
\usepackage{epsfig}
\usepackage{color}
\usepackage{float}
\usepackage[normalem]{ulem}
\usepackage{comment}
\usepackage{times}
\usepackage{epstopdf}
\usepackage{amssymb}
\usepackage{pdflscape}
\usepackage{latexsym}

\usepackage[OT2,T1]{fontenc}
\DeclareSymbolFont{cyrletters}{OT2}{wncyr}{m}{n}
\DeclareMathSymbol{\Sha}{\mathalpha}{cyrletters}{"58}
\DeclareMathSizes{12}{20}{4}{10}

\def\del{\delta}
\def\tdel{\widetilde{\delta}}

\newcommand{\be}{\begin{equation}}
\newcommand{\ee}{\end{equation}}
\newcommand{\bdm}{\begin{displaymath}}
\newcommand{\edm}{\end{displaymath}}
\newcommand{\bea}{\begin{eqnarray}}
\newcommand{\eea}{\end{eqnarray}}
\newcommand{\nn}{\nonumber}
\newcommand{\kv}{{\bf k}}

\newcommand{\qv}{{\bf q}}

\newcommand{\rv}{{\bf r}}

\newcommand{\xv}{{\bf x}}

\newcommand{\nv}{{\bf n}}

\newcommand{\kN}{{k_{\rm Nyq}}}

\newcommand{\de}{{\rm d}}

\newcommand{\eq}[1]{eq.\,(\ref{#1})}

\def\Oc{{\mathcal O}}
\def\Mpc{\, h^{-1} \, {\rm Mpc}}

\def\kMpc{\, h \, {\rm Mpc}^{-1}}

\def\powmes{\texttt{POWMES}}

\usepackage[pdftex]{hyperref}

\title[Accurate Estimators of Correlation Functions in Fourier Space]{Accurate Estimators of Correlation Functions in Fourier Space}
\author[E. Sefusatti, {\em et al.}]
{E. Sefusatti$^{1,2}$\footnotemark[1]\thanks{E-mail:sefusatti@oats.inaf.it (ES)},
 M. Crocce$^{3}$\footnotemark[1]\thanks{E-mail:crocce@ice.cat (MC)},
R. Scoccimarro$^{4}$\thanks{E-mail:rs123@nyu.edu (RS)},
and
H. M. P. Couchman$^{5}$\thanks{E-mail:couchman@physics.mcmaster.ca (HC)}
\vspace{1mm}\\
$^{1}$INAF - Osservatorio Astronomico di Brera, via E. Bianchi 46, 23807 Merate (LC), Italy\\
$^{2}$INFN - Sezione di Padova, via Marzolo 8, 35131 Padova, Italy\\
$^{3}$Institut de Ci\`encies de l'Espai, IEEC-CSIC, Campus UAB, Carrer
de Can Magrans, s/n, 08193 Bellaterra, Barcelona, Spain\\
$^{4}$Center for Cosmology and Particle Physics, Department of Physics, New York University, NY 10003, New York, USA\\
$^{5}$Department of Physics and Astronomy, McMaster University, Hamilton, ON L8S 4M1, Canada}

\begin{document}

\date{\today}

\maketitle

\label{firstpage}

\begin{abstract}

Efficient estimators of Fourier-space statistics for large number of objects rely on Fast Fourier Transforms (FFTs), which are affected by aliasing from unresolved small scale modes due to the finite FFT grid. Aliasing takes the form of a sum over images, each of them corresponding to the  Fourier content displaced by increasing multiples of the sampling frequency of the grid. 
These spurious contributions limit the accuracy in the estimation of Fourier-space statistics, and are typically ameliorated by simultaneously increasing grid size and discarding high-frequency modes. This results in inefficient estimates for e.g. the power spectrum when desired systematic biases are well under per-cent level. 
 We show that using interlaced grids removes odd images, which include the dominant contribution to aliasing. In addition, we discuss the choice of interpolation kernel used to define density perturbations on the FFT grid and demonstrate that using higher-order interpolation kernels than the standard Cloud in Cell algorithm results in significant reduction of the remaining images.  We show that combining fourth-order interpolation with interlacing gives very accurate Fourier amplitudes and phases of density perturbations. This results in power spectrum and bispectrum estimates that have systematic biases below $0.01\%$ all the way to the Nyquist frequency of the grid, thus  maximizing the use of unbiased Fourier coefficients for a given grid size and greatly reducing systematics for  applications to large cosmological  data sets.
  \end{abstract}


\section{Introduction}

A great effort is currently directed towards accurate measurements of the power spectrum of the galaxy distribution as a test of the cosmological model \citep[see, e.g.][]{LaureijsEtal2011, LeviEtal2013}. At the same time, theoretical predictions for the matter and galaxy power spectrum, which play a key role in the interpretation of observational data, are ultimately validated by measurements of such quantities in numerical simulations. Due to computational efficiency, Fourier-space statistics in simulations and galaxy surveys are usually computed using Fast Fourier Transforms (FFTs), which require building estimates of density or galaxy perturbations on a regular grid from the input data objects ($N$-body particles or galaxies). These Fourier coefficients are then used to calculate correlation functions in Fourier space such as the power spectrum or bispectrum.     

However, introducing a discrete sampling as the FFT grid comes with limitations that can only be avoided under special circumstances. If the underlying spectrum of perturbations is bandwidth limited (vanishes above a cutoff frequency $k_{\rm max}$), then the sampling theorem guarantees that grid estimates are lossless as long as the sampling frequency of the grid ($k_s = 2\pi/H$, where $H$ is the grid spacing) is at least twice the cutoff frequency, $k_s \ge 2 k_{\rm max}$. This just says that the grid is fine enough so that every Fourier component in the signal is sampled at least twice per period, which intuitively makes sense (at least a peak and a trough is needed to resolve an oscillation). 

Unfortunately the conditions of the sampling theorem are not satisfied by cosmological perturbations, which have significant Fourier content at all frequencies. As a result of this, unresolved small-scale modes when sampled by the grid are mistakenly identified as modes supported by the grid, which results in spurious  contamination of the FFT-determined Fourier coefficients that directly translates into a systematic error in e.g. the power spectrum and bispectrum. This contamination is known as {\em aliasing}: it takes the form of a sum over images, each of them corresponding to the  Fourier content displaced by increasing multiples of the sampling frequency of the grid $k_s$, as we shall discuss in detail below. 

Aliasing can be removed entirely if the signal is first low-pass filtered so that only frequencies supported by the grid remain before sampling by the grid, effectively making the modified signal bandwidth limited. Although such process is not lossless (high frequencies are missing), the determination of the supported Fourier modes is free of systematics. Low-pass filtering, however, results in interpolating kernels that are very non-local in real space (a given object contributes to essentially all grid points), and thus the definition of perturbations on the grid becomes too computationally expensive to be useful despite the speedup of FFTs. The goal to minimize aliasing is therefore to design an interpolating kernel that is fairly local while at the same time acts as much as possible as a low-pass filter (i.e. it is fairly smooth). In this paper we present an algorithm that achieves this for realistic spectra of cosmological perturbations. By working at the level of the Fourier modes, it simultaneously  benefits all Fourier-space statistics.  

The issue of aliasing in the measurements of the power spectrum has been addressed, in the specific context of cosmological studies, by \citet{Jing2005, CuiEtal2008, YangEtal2009, ColombiEtal2009} and \citet{JascheKitauraEnsslin2009}. In particular, \citet{Jing2005} derives the full expression, including the aliasing component, for the power spectrum measured from interpolation on a grid with a generic mass assignment scheme and proposes an iterative procedure to reduce the aliasing contribution based on the exact knowledge of the window function in Fourier space corresponding to such scheme.  \citet{CuiEtal2008} and  \citet{YangEtal2009} do not attempt to correct for the error introduced by aliasing but search instead for an optimal assignment scheme able to reduce the effect. They identify for this purpose the scaling function of specific Daubechies wavelets and claim, as a result, a systematic error below the 2\% level for wavenumbers $k<0.7 \kN$, $\kN\equiv k_s/2$ being the Nyquist frequency of the grid.   
A different approach is explored by \citet{ColombiEtal2009} who consider a Fourier-Taylor expansion of the expression corresponding to a direct summation of the particle contribution to the density field in Fourier space. At the lowest order the procedure, implemented in the {\powmes} public code, corresponds to a Nearest Grid Point (NGP)  assignment scheme but including higher-order corrections it quickly converges to an unbiased estimator of the power spectrum, albeit at significant computational cost. Finally, \citet{JascheKitauraEnsslin2009} discuss a method based on ``supersampling'' that inevitably increases memory requirements (typically by almost an order of magnitude). 

In this paper we revisit a technique based on interlacing two density grids that, while requiring only two times the number of density evaluations w.r.t. the standard case, can significantly reduce the aliasing contribution. The method is discussed in \citet{HockneyEastwood1981}, and has been  implemented in the AP$^3$M $N$-body code for the mesh force calculation in \citet{Couchman1991} \citep[see also][]{Couchman1999}. Here we show how it can be applied to the evaluation of the density field in Fourier space from point distributions. In particular we quantify the relative effect of different mass assignment schemes on the aliasing contributions to the power spectrum estimation in two typical situations of cosmological interest: the nonlinear matter distribution at low redshift from a small N-body simulation and the matter distribution obtained in the Zeldovich Approximation (ZA) at high redshift.  We proceed to illustrate the interlacing technique and we compare our results to Fourier coefficient evaluations by direct summation over objects (which are alias-free) and to the output of the {\powmes} public code. 

The paper is organized as follows. Section \ref{sec:aliasing} introduces and describes the problem of aliasing in mathematical terms, with specific attention to its interplay with the order of the mass assignment scheme adopted. Section \ref{sec:interlacing} presents the interlacing technique and its effect on different power spectrum  measurements. In Section \ref{sec:higherorder} we quantify the contribution of aliasing with and without interlacing to higher order correlation functions such as the bispectrum. We present our conclusions in Section \ref{sec:conclusions}.

\section{Density estimation and aliasing}
\label{sec:aliasing}

\subsection{Correlation functions in Fourier space}

We focus our attention on continuous fields defined over a finite, cubic volume $V=L^3$ which we will assume equivalent to the case of infinite volume but periodic with period $L$ along each of the three spatial directions (the extension to the general case of a generic box of different periods while trivial, would require a significantly more cumbersome notation). 

The matter density $\rho(\xv)$ can be described in terms of the dimensionless overdensity
\be
\del(\xv)\equiv\frac{\rho(\xv)}{\bar{\rho}}-1\,,
\ee
where $\bar{\rho}$ is the mean density in the volume $V$. Its Fourier space counterpart is given by
\be
\del(\kv)  \equiv  \int_V \frac{d^3x}{(2\pi)^3}\, e^{-i\kv\cdot\xv}\,\del(\xv)\,,
\ee
while its inverse is defined in terms of the series
\be
\del(\xv)  \equiv  k_f^3\, \sum_{\kv}\,e^{i\kv\cdot\xv}\,f(\kv)\,,
\ee
since $\kv$ is discretised in multiples of the fundamental frequency $k_{f}=2\pi/L$ in each spatial direction. Notice that we will always explicitly show the argument of each function in order to distinguish a real-space object from its Fourier transform.\footnote{In this convention for the Fourier transform,  the Dirac delta in real space and the Kronecker delta in Fourier space are given, respectively by the following representations
\bdm
\delta_D(\xv)=\frac1V\sum_{\kv} \,e^{i\,\kv\cdot\xv}\quad {\rm and} \quad \del^K_{\kv}=\frac1V\int d^3x\,e^{-i\,\kv\cdot\xv}\,.
\edm
}

The power spectrum $P(k)$ is then defined as the two-point function of $\del(\kv)$, that is
\be\label{eq:ps}
\langle \del(\kv_1)\,\del(\kv_2)\rangle \equiv \frac{\delta_{\kv_{12}}^K}{k_f^3}\,P(k_1)\,,
\ee
with $\langle\, \dots\rangle$ representing the ensemble average, $\kv_{i_1,\dots,i_n}\equiv\kv_{i_1}+\dots+\kv_{i_n}$ and $\delta^K_\kv$ being a Kronecker delta defined to be equal to 1 for a vanishing argument $\kv=0$ and zero otherwise  . Similarly, the bispectrum $B(k_1,k_2,k_3)$ is defined as 
\be
\langle \del(\kv_1)\,\del(\kv_2)\,\del(\kv_3)\rangle \equiv \frac{\del_{\kv_{123}}^K}{k_f^3}\,\,B(k_1,k_2,k_3)\,.
\ee
In this convention the continuous limit is recovered for $V\rightarrow \infty$ or $k_f\rightarrow 0$  with $\del^K_\kv/k_f^3\rightarrow \del_D(\kv)$, the latter being a Dirac delta function.

\subsection{Direct summation}

Theoretical predictions for the correlation functions of the matter density, or of the galaxy number density, often assume such quantities to be continuous random fields. However, in practical applications such as the analysis of $N$-body simulations or galaxy surveys they are given in terms of a finite number $N_{\rm P}$ of objects with positions $\left\{\xv_i\right\}$ for $i=1,\dots,N_{\rm P}$. In this case we can write the density as
\be\label{eq:rhosum}
\rho(\xv)=\sum_{i=1}^{N_{\rm P}}\,m\,\delta_D(\xv-\xv_i)\,,
\ee
$m$ being the particle mass which we assume here to be the same for all particles, for simplicity. It follows that 
\be\label{eq:deltasum}
\del(\xv) =\frac{1}{\bar{n}}\sum_{i=1}^{N_{\rm P}}\,\delta_D(\xv-\xv_i)-1\,,
\ee
$\bar{n}\equiv N_{\rm P}/V$ being the particle density. The overdensity in Fourier space can be obtained by direct summation as the Fourier transform of the equation above, that is 
\be
\del(\kv)  
 =  \frac{1}{(2\pi)^3}\,\frac1{\bar{n}}\,\sum_{i=1}^{N_{\rm P}} e^{-i\,\kv\cdot\xv_i}- \frac{\del^K_{\kv}}{k_f^3}\,.
\label{eq:deltaDS}
\ee

The Fourier-space 2-point function defining the power spectrum is now given by
\be\label{eq:ds2pcf}
\langle\,\del(\kv_1)\,\del(\kv_2)\,\rangle=\frac{\del_{\kv_{12}}^K}{k_f^3}\,\left[P(k_1)+\frac{1}{(2\pi)^3\,\bar{n}}\right]\,,
\ee
where the second term on the r.h.s. represents the shot-noise contribution due to the self-correlation of individual particles  \citep[see, for instance,][]{Peebles1980, Jing2005}. 
The direct summation estimator for the power spectrum of a particle distribution is then 
\be
\hat{P}(k)\equiv k_f^3\,|\del(\kv)|^2= k_f^3\left[|\del(\kv)|^2-\frac{1}{N_{\rm P}}\right] \,,
\ee
and the true power spectrum is recovered as $P(k)=\langle\hat{P}(k)\rangle$. 

Direct summation is clearly quite a demanding approach, particularly in the case of $N$-body simulations with a very large number of particles, since the evaluation of the Fourier-space density scales as $N_{\rm P}\times\,N_G^3$, $N_G^3$ being the number of wavenumbers $\kv$ required. While impractical, it is however an exact alias-free determination of Fourier coefficients. Therefore we shall use direct summation estimates below as a benchmark to which FFT methods are compared.

\subsection{Mass Assignment and Aliasing}

A more efficient method, taking advantage of the FFT algorithm, requires first the interpolation of the density field on a regular grid in position space. This fixes the largest wavenumber accessible to the Nyquist frequency, $\kN=\pi N_G/L=\pi/H$, $N_G$ being now the linear size of the grid and $H=L/N_G$ therefore being the grid spacing. The price to pay is the emergence of aliasing in the Fourier transform of the density grid and therefore a systematic error that cannot be easily and completely removed from the measurements of the correlation functions in Fourier space. We will loosely revisit here the derivation by \citet{Jing2005} of the theoretical expression for the power spectrum estimator based on grid interpolation and the FFT, since it will be useful to describe the interlacing method discussed in section~\ref{sec:interlacing}. 

Let us assume $\xv_j^G$ with $j=1$, ... $N_G^3$ to be a regular grid of points, for simplicity, linearly spaced by $H$ in all directions.  The interpolation of a continuous function $f(\xv)$ over the grid can be given by a simple integration over the cell of volume $H^3$ surrounding each point, so that
\bea
f(\xv_j^G) &\!\! =\!\! & \frac1{H^3}\,\int_{\xv_j^G}d^3 x\,f(\xv)\nn \\
& \!\!\equiv\!\! & \frac1{H^3}\,\int_{x_j^G-H/2}^{x_j^G+H/2}\!\!\!\! dx\,\int_{y_j^G-H/2}^{y_j^G+H/2}\!\!\!\!dy\,\int_{z_j^G-H/2}^{z_j^G+H/2}\!\!\!\!dz\,f(\xv)\,.
\eea
In the case of the density defined by a particle distribution as in \eq{eq:rhosum}, each particle will provide a contribution only to the cell encompassing the particle position. As we will see, this simple procedure, known as the Nearest Grid Point (NGP) assignment scheme, is not the optimal one. To consider more general schemes, it is convenient to assign a ``shape'' to each particle defined by a function $S(\xv)$ which is symmetric, positive defined, separable as $S(\xv)=S_{\rm 1D}(x)\,S_{\rm 1D}(y)\,S_{\rm 1D}(z)$, and normalised such that $\int d^3x  S(\xv)/{(2\pi)^3}=1$. Separability makes computations efficient, since the resulting scheme will only need 1D distances in each direction, as opposed to e.g. spherically symmetric shapes which require radial distances to be computed. Similarly, correcting for  the interpolation window in Fourier space is straightforward when separability is assumed. 

The interpolation over the grid corresponds then to the evaluation of the (continuous) function
\be\label{eq:tdel}
\tdel(\xv) \equiv \int \frac{d^3x'}{(2\pi)^3}\,W(\xv-\xv')\,\del(\xv')
\ee
at the grid points $\xv^G_j$, leading to
\be
\tdel(\xv^G_j)  =  \frac1{{(2\pi)^3}\bar{n}}\sum_{i=1}^{N_P}\,W(\xv^G_j-\xv_i)-1\,,
\label{eq:delgrid}
\ee
where the weight function $W(\xv)$ is simply the integral over the cell volume of the shape function $S(\xv)$ given by
\be
W(\xv_j^G-\xv_i)  =\frac1{H^3}\,\int_{\xv_j^G}d^3 x\,S(\xv-\xv_i)\,.
\ee

In addition to the Nearest Grid Point, assignment schemes commonly adopted are the Cloud In Cell and Triangular Shape Cloud scheme, representing the lowest order piecewise polynomial function (also known as B-splines). They correspond, respectively, to first, second and third order referring to the number $p$ of grid points, per dimension, to which each particle is assigned. The relative piecewise polynomial functions $W^{(p)}$ are of order less or equal to $(p-1)$, and correspond simply to convolving a top-hat function with itself $(p-1)$ times. The weights are then given explicitly by $W(\xv)=W^{(p)}(x/H)\,W^{(p)}(y/H)\,W^{(p)}(z/H)$, where for each different scheme we have the one-dimensional functions:
\begin{itemize}[leftmargin=10pt]
\item Nearest Grid Point (NGP)
\be
W^{(1)}(s)=\left\{
\begin{array}{ll}
1 & {\rm for}~ |s|<\frac12 \\
0 & {\rm otherwise}
\end{array}
\right.
\ee
\item Cloud-In-Cell (CIC)
\be
W^{(2)}(s)=\left\{
\begin{array}{ll}
1-|s| & {\rm for}~ |s|<\frac12 \\
0 & {\rm otherwise}
\end{array}
\right.
\ee
\item Triangular Shaped Cloud (TSC)
\be
W^{(3)}(s)=\left\{
\begin{array}{ll}
\frac34-s^2 & {\rm for}~ |s|<\frac12 \\
\frac12\,\left(\frac32-|s|\right)^2 & {\rm for}~ \frac12\le |s|<\frac32 \\
0 & {\rm otherwise}
\end{array}
\right.
\ee
\end{itemize}
where $s=(x_j^G-x)/H$.
Clearly, schemes of order higher than TSC can be considered \citep[as it is often the case in plasma physics applications, see, e.g.][]{HaugbolleEtal2013}. Here we will consider only the next, fourth-order interpolation scheme, which we will refer to as Piecewise Cubic Spline \citep[see, e.g.][for generic, higher-order B-spline interpolation]{ChaniotisPoulikakos2004}. The corresponding weights are given by
\begin{itemize}[leftmargin=10pt]
\item Piecewise Cubic Spline (PCS)
\be
W^{(4)}(s)=\left\{
\begin{array}{ll}
\! \!\!\frac16\,\left(4-6\,s^2+3\,|s|^3\right) &\! \!\!{\rm for}~ 0\le |s|< 1 \\
\! \!\!\frac16\,\left(2-|s|\right)^3 &\!\!\!  {\rm for}~ 1 \le |s|< 2 \\
\! \!\!0 & {\rm otherwise}
\end{array}
\right.
\ee
\end{itemize}
    
\begin{figure*}
\includegraphics[width=0.88\textwidth]{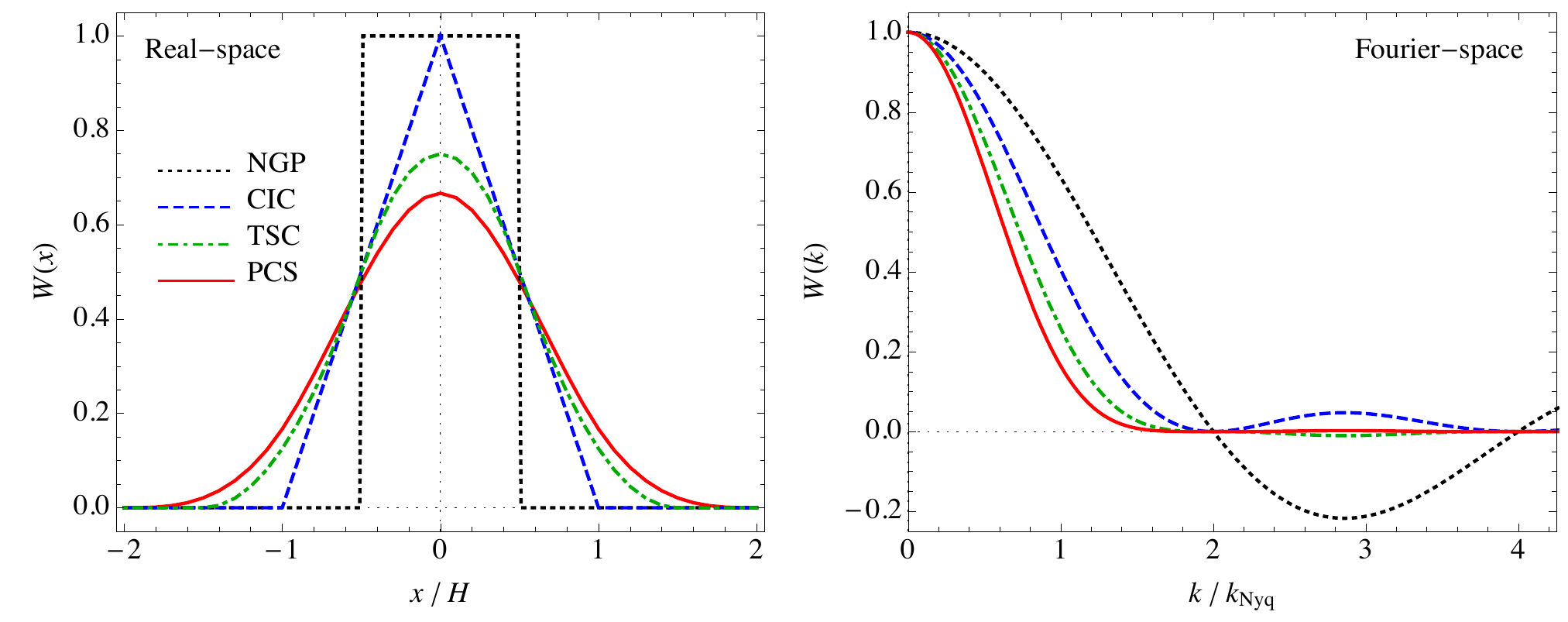}
 \caption{{\em Left panel}: one-dimensional, real-space window
   functions $W^{(p)}(x)$ of the four mass assignment schemes described in the text. Different curves show Nearest Grid Point (NGP, {\em
     dotted, black}), Cloud In Cell (CIC, {\em dashed, blue}),
   Triangular Shaped Cloud (TSC, {\em dotdashed, green}) and Piecewise Cubic Spline 
   interpolation (PCS, {\em continuous, red}). {\em Right panel}: the same
   window functions in Fourier space, $W^{(p)}(k)$.}
 \label{fig:mas}
\end{figure*}

In the left panel of Fig.~\ref{fig:mas} we show the one-dimensional window functions in real space $W^{(p)}(x)$ where it is evident how the support of the weight function grows with the interpolation order. 

The sampling in real space given by \eq{eq:delgrid} can be mathematically described in terms of the sampling function $\Sha(\xv)$ \citep[see, {\em e.g.}][]{HockneyEastwood1981}. This is defined as
\be
\Sha(\xv)=\sum_{\nv}\delta_D(\xv-\nv)\,,
\ee 
with $\nv$ an integer vector. It is easy to see that the Fourier transform of $\widetilde{\del}(\xv_j^G)$ is equal to the Fourier transform of the continuous field given by the product $\Sha(\xv/H)\,\widetilde{\del}(\xv)$ which is different from zero only for $\xv=\xv^G_j$ and equals $\widetilde{\del}(\xv_j^G)$ at those points when integrated over.  Applying the convolution theorem we obtain
\be\label{eq:tdelG1}
\tdel^{\rm G}(\kv)=k_f^3\,\sum_{\kv'}\,{\Sha}(\kv')\,\tdel(\kv-\kv')\,,
\ee
where $\tdel(\kv)$ is the Fourier transform of the convolved, continuous density field defined in \eq{eq:tdel}, while $\tdel^G(\kv)$, in addition, accounts for the discreteness of the grid assignment.  

We notice now that the Fourier transform of the sampling function is a sampling function in $k$-space. Since we are dealing with periodic functions in position space the Delta functions are replaced by Kronecker deltas as
\be\label{eq:FTsha}
\Sha(\kv)=\frac{1}{k_f^3}\,\sum_n\,\delta^K_{\kv-\nv \, k_s}\,,
\ee 
where the sampling frequency of the grid $k_s$ is given by
\be\label{eq:ks}
k_s \equiv {2\pi \over H}
\ee
so that we obtain
\be\label{eq:tdelG2}
\tdel^{\rm G}(\kv)  =  \sum_{\nv}\,\tdel\left(\kv-\nv\, k_s\right) \,,
 \ee
where clearly $\tdel^{\rm G}$ is $k_s$-periodic and we are thus interested in frequencies  $|\kv| \leq k_s/2 = k_{\rm Nyq}=\pi/H=\pi\,N_G/L$, with $k_{\rm Nyq}$ the Nyquist frequency of the grid. In this expression, all terms with $\nv\ne 0$ in the sum represent the aliasing contribution to the estimate $\tdel^{\rm G}(\kv)$ of the density $\tdel(\kv)$, which, we should keep in mind, includes the effect of the  interpolation function $W(\xv)$. Such effects can be  removed from the leading contribution in the sum above, simply dividing by the FT of the window function itself, $W(\kv)$.  The {\em corrected}, grid-based density contrast would then be given by
\be\label{eq:windowcorrection}
\del^{\rm G}(\kv)  \equiv  \frac{\tdel^{\rm G}(\kv)}{W(\kv)}
   =  \frac1{W(\kv)}\,\sum_{\nv}\,\tdel\left(\kv-\nv\, k_s\right) \,,
 \ee
and introducing the ``corrected'' window function
\be
w_\nv(\kv)\equiv\frac{W(\kv-\nv\, k_s)}{W(\kv)}\,,
\ee
defined in such a way that $w_0(\kv)=1$,  we obtain
\be\label{eq:this}
\del^{\rm G}(\kv) 
   =  \del\left(\kv\right) +\sum_{\nv\ne{\bf 0}}\,w_\nv(\kv)\ \del\left(\kv-\nv\, k_s\right) \,,
 \ee
where indeed the $\nv=0$ term provides the desired estimation of the density $\del(\kv)$, while all other terms correspond to spurious,
aliasing contributions. They correspond to ``images" of the Fourier content $\del(\kv)$ centred at increased multiples of the sampling frequency of the grid $k_s$. For the frequencies of interest, $|\kv| \leq k_s/2 = k_{\rm Nyq}$, the images correspond to small-scale modes not supported by the grid that masquerade as modes of  the frequency range we are interested in. These unwanted contributions have an amplitude that depends not only on the small-scale power but also on the  weight given by the corrected window function $w_\nv$. 

Our main goal here is therefore to minimise $w_\nv$ subject to the constraints of computational speed. Ideally, as pointed out in the introduction, to remove aliasing one would choose an interpolation function $W(k)$ which is a low-pass filter for modes of frequency less than $k_s$, which would automatically make the $w_\nv$ vanish. The drawback of this choice is that the real-space kernel is then very non-local and  that makes inefficient the interpolation to the grid, since each object would contribute to a large  number of grid points. This, in effect, would constitute the bottleneck of the computational cost, and would negate the whole point of using FFTs as opposed to direct summation. The objective is then to find interpolation kernels that are sufficiently local in both real and Fourier space.

In order to highlight these aspects, on the right panel of Fig.~\ref{fig:mas} we show the (one-dimensional) window functions in Fourier space $W(k)$. For the B-spline windows considered above these are simply given by
\be
W^{(p)}(k)=\left[\frac{\sin (kH/2)}{(kH/2)}\right]^p=\left[\frac{\sin (\pi \, k/k_s)}{(\pi\, k/k_s)}\right]^p\,,
\label{eq:intkerFT}
\ee
$p=1$ to 4 being the interpolation order corresponding to NGP, CIC, TSC and PCS. Again, Eq.~(\ref{eq:intkerFT}) makes clear that the different interpolation schemes correspond to convolutions of $(p-1)$ top-hat functions in 1D. For the 3D case, due to the separability assumption, one simply multiplies three such objects for each dimension, which is quite efficient. It is clear from  Fig.~\ref{fig:mas} that the PCS kernel satisfies the opposite requirements of compactness in real and Fourier space rather well, although there is of course a tradeoff. The Fourier-space compactness means that the amplitude of the unwanted images $w_\nv$ will be highly suppressed, this comes at the expense of a slightly more expensive interpolation to the grid as the real-space kernel is correspondingly broader. We will examine the impact of these choices for the power spectrum and bispectrum below.

As mentioned above, any aliasing correction to the density field will inevitably affect the estimation of any correlation function. In the first place, the two-point function of the interpolated, corrected density field $\del^{\rm G}(\kv)$, will be given by
\bea
~~~\langle\,\del^{\rm G}(\kv_1)\,\del^{\rm G}(\kv_2)\,\rangle 
& = & \sum_{\nv_1,\,\nv_2}\,{w_{\nv_1}(\kv_1)\,w_{\nv_2}(\kv_2)} \nn\\
& & \times\,\langle\,\del(\qv_1)\,\del(\qv_2)\,\rangle\,,
\eea
where $\qv_i=\kv_i-\nv_i\, k_s$ and where the expectation value on the r.h.s. coincides with the one of \eq{eq:ds2pcf}. 

The relation between the power spectrum estimated from the grid-based density $\del^{\rm G}$ and the desired power spectrum $P(k)$ of the original distribution is then given by
\bea\label{eq:Pknoint}
~~~P^{\rm G}(k)\ &\!\!\!\!\!\equiv\!\!\!\!\! & k_f^3\,\langle\,|\del^{\rm G}(\kv)|^2\,\rangle \nonumber\\
&\!\!\!\!\! = \!\!\!\!\!& \sum_{\nv}\,\left|w_{\nv}(\kv)\right|^2  \,P_{tot}\left(|\kv\!-\!\nv\, k_s|\right)\,\!\!,
\eea
where $P_{tot}(k)\,\equiv\, P(k)+1/[(2\pi)^3\,\bar{n}]$, an expression first derived by \citet{Jing2005}, here corrected for the window effects at leading order in the aliasing expansion.

In the case of the bispectrum $B^{\rm G}(k_1,k_2,k_3)$, we obtain the similar expression,
\bea
B^{\rm G}(k_1,k_2,k_3)
& \!\!\!\!= \!\!\!\!& \sum_{\nv_1,\,\nv_2,\,\nv_3}\,w_{\nv_1}(\kv_1)\,w_{\nv_2}(\kv_2)\,w_{\nv_3}(\kv_3)\nonumber\\
& & \times\delta^K_{\qv_{123}} \,B_{tot}(q_1,q_2,q_3)\nn\\
& \!\!\!\!= \!\!\!\!& \sum_{\nv_1,\,\nv_2}\,w_{\nv_1}(\kv_1)\,w_{\nv_2}(\kv_2)\,w_{-\nv_{12}}(-\kv_{12})\nonumber\\
& & \times \,B_{tot}(\kv_1-\nv_1\,k_s,\kv_2-\nv_2\,k_s)\,,
\eea
where
\bea
B_{tot} & \equiv & B(k_1,k_2,k_3)
+\frac1{(2\pi)^3\,\bar{n}}\left[P(k_1)+P(k_2)+P(k_3) \right]
\nn\\ & & 
+\frac1{(2\pi)^6\,\bar{n}^2}\,,
\eea
 and where, again, only the leading term for $\nv_1=\nv_2=\nv_3=0$ corresponds to the desired estimated bispectrum (plus shot-noise), while all the others provide unwanted aliasing contributions.

\citet{Jing2005} proposes a procedure to correct for the aliasing contribution to the power spectrum by means of a function that requires, in principle, a priori knowledge of the target power spectrum itself. Since, however, such knowledge is relevant only for values close to the Nyquist frequency, a power-law approximation of the power spectrum, determined by an iterative scheme, is employed instead. While this could provide a reasonable and effective solution to the problem of aliasing in power spectrum estimation, the procedure cannot be easily extended to the bispectrum, where one would need a model to describe the scale and triangle-shape dependence of high-$k$ modes to effectively handle such corrections. 

We will explore, in the rest of the paper, a more flexible  alternative.

\section{Aliasing reduction by interlacing}
\label{sec:interlacing}

\subsection{Interlacing}

A method to partially correct for aliasing based on the interlacing of two grid is discussed in \citet{HockneyEastwood1981}, and its implementation in the AP$^3$M $N$-body code by \citet{Couchman1991} for the mesh force calculation is mentioned in \citet{Couchman1999}. 

We have seen that, for a generic mass assignment scheme, the interpolated density field, denoted below as $\tdel_1^{\rm G}(\kv)$, can be written as in \eq{eq:tdelG2}, which we reproduce here for convenience
\bea
\tdel_1^{\rm G}(\kv) & = & \int_V \frac{d^3x}{(2\pi)^3}\, e^{-i\,\kv\cdot\,\xv}\,\Sha\left(\frac{\xv}{H}\right)\,\tdel(\xv) \nonumber \\
& = & \sum_{\nv}\,\tdel\left(\kv-\nv\, k_s\right) \,.
\eea
An additional, real-space interpolation on a grid shifted by the distance $H/2$ in all spatial directions, with respect to the one above can be written as
\be
\tdel_2^{\rm G}(\kv)  = \int_V \frac{d^3x}{(2\pi)^3}\, e^{-i\,\kv\cdot\,\xv}\,\Sha\left(\frac{\xv}{H}+\frac12 \right)\,\tdel(\xv)\,.
\ee
Again, since this represents the FT of a product, it can be written as a convolution of Fourier series, where, this time, the transform of the sampling function, \eq{eq:FTsha}, is obtained taking advantage of the shift theorem so that 
\bea
~~~\tdel_2^{\rm G}(\kv) & \!\!=\!\! & k_f^3\,\sum_{\kv'}\,e^{-i\,(k'_x+k'_y+k'_z)\,H/2}\,\Sha(\kv')\,\tdel(\kv-\kv') \nonumber\\
 & \!\!=\!\! & \sum_{\nv}\,e^{-\pi\,i\,(n_x+n_y+n_z)}\,\tdel\left(\kv-\nv\, k_s\right) \nonumber\\
 &\!\! =\!\! & \sum_{\nv}\,(-1)^{n_x+n_y+n_z}\,\tdel\left(\kv-\nv\, k_s\right) \,.
\eea
Considering now the linear combination of the two, one obtains
\bea\label{eq:intdelta}
~~\tdel^{\rm G}(\kv) &\!\!\!\! =\!\!\!\! & \frac12\left[\tdel_1^{\rm G}(\kv) +\tdel_2^{\rm G}(\kv) \right]   \nonumber\\
& \!\!\!\!=\!\!\!\! & \sum_{\nv}\,\theta_\nv\,\tdel\left(\kv-\nv\, k_s\right),
\eea
where we defined
\bea
\theta_\nv & \equiv & \frac12 \left[1\!+\!(-1)^{n_x+n_y+n_z}\right] \nn\\
&  = & \left\{\begin{array}{ll} 1 & {\rm for} ~n_x+n_y+n_z ~{\rm even}\\
0 & {\rm for} ~n_x+n_y+n_z ~{\rm odd}
\end{array}
\right.
\eea
In the expression of \eq{eq:intdelta} for the interlaced density all aliasing contributions (images) corresponding to odd values of the sum $n_x+n_y+n_z$, including, in particular,  the largest ones for $|\nv|=1$ are removed. Clearly, the density field thus obtained should still be corrected for the effects of the window function as in \eq{eq:windowcorrection}, so that $\del^G(\kv)=\tdel^G(\kv)/W(\kv)$. In principle, the scheme could be extended to the combination of more than two interpolated density grids to remove the leading aliasing contributions at the next order. 

The power spectrum of the interlaced density field, corrected for the window function, can now be written as
\bea
~~~P^{\rm G}(k)\ &\!\!\!\!\! = \!\!\!\!\! & \sum_{\nv}\,\theta_\nv\,\left|w_{\nv}(\kv)\right|^2 \,P_{tot}\left(|\kv\!-\!\nv\, k_s|\right)\,,
\eea
while the expression for the power spectrum of the (not interlaced) field of \eq{eq:Pknoint} is then recovered substituting $\theta_\nv$ with 1 in the equation above.  

We can write the residual aliasing contribution as
\bea
\Delta P^G(k) & \equiv & P^G(k) - P_{tot}(k)\nonumber\\
& \!\!\!\!=\!\!\!\! & \sum_{\nv\ne 0}\,\theta_{\nv}\,|w_{\nv}(\kv)|^2\,P_{tot}(\kv-\nv\, k_s)\,.
\eea
Assuming that, for all $k>\kN$, the power spectrum has an upper bound $P_{\rm max}\ge P(k)$, the residual is, in turn, bounded as 
\bea
\Delta P^G(k) & \le & \left[P_{\rm max}+\frac1{(2\pi)^3\,\bar{n}}\right]\,F_{\rm res}(\kv)\,,
\eea
where we introduce the residual factor $F_{\rm res}(\kv)$ on the r.h.s., defined as
\bea\label{eq:res}
F_{\rm res}(\kv) & \equiv& \sum_{\nv\ne 0}\,\theta_{\nv}\,|w_{2\nv}(\kv)|^2\,.
\eea
For mass assignment schemes corresponding to B-splines as those considered before, it is possible to evaluate this residual factor explicitly. For the interpolation order $p$, we have 
\bea\label{eq:Fres} 
F_{\rm res}(\xv)
 & = &   
\sum_{\nv\ne 0}\,\theta_{\nv}\,\prod_{i=1}^3 \, \left(\frac{x_i}{x_i-n_i}\right)^{2\,p}\,,
\eea
where $\xv\equiv \kv/k_s$ and the product runs overs the three spatial dimensions. With no interlacing we obtain, more simply
\bea\label{eq:Fres} 
F_{\rm res}(\xv)
 & = &   
\sum_{\nv\ne 0}\,\prod_{i=1}^3 \, \left(\frac{x_i}{x_i-n_i}\right)^{2\,p}\,.
\eea
We note that in these expressions a relatively small number of terms are enough to recover a reasonable accuracy.

\begin{figure}
\includegraphics[width=.44\textwidth]{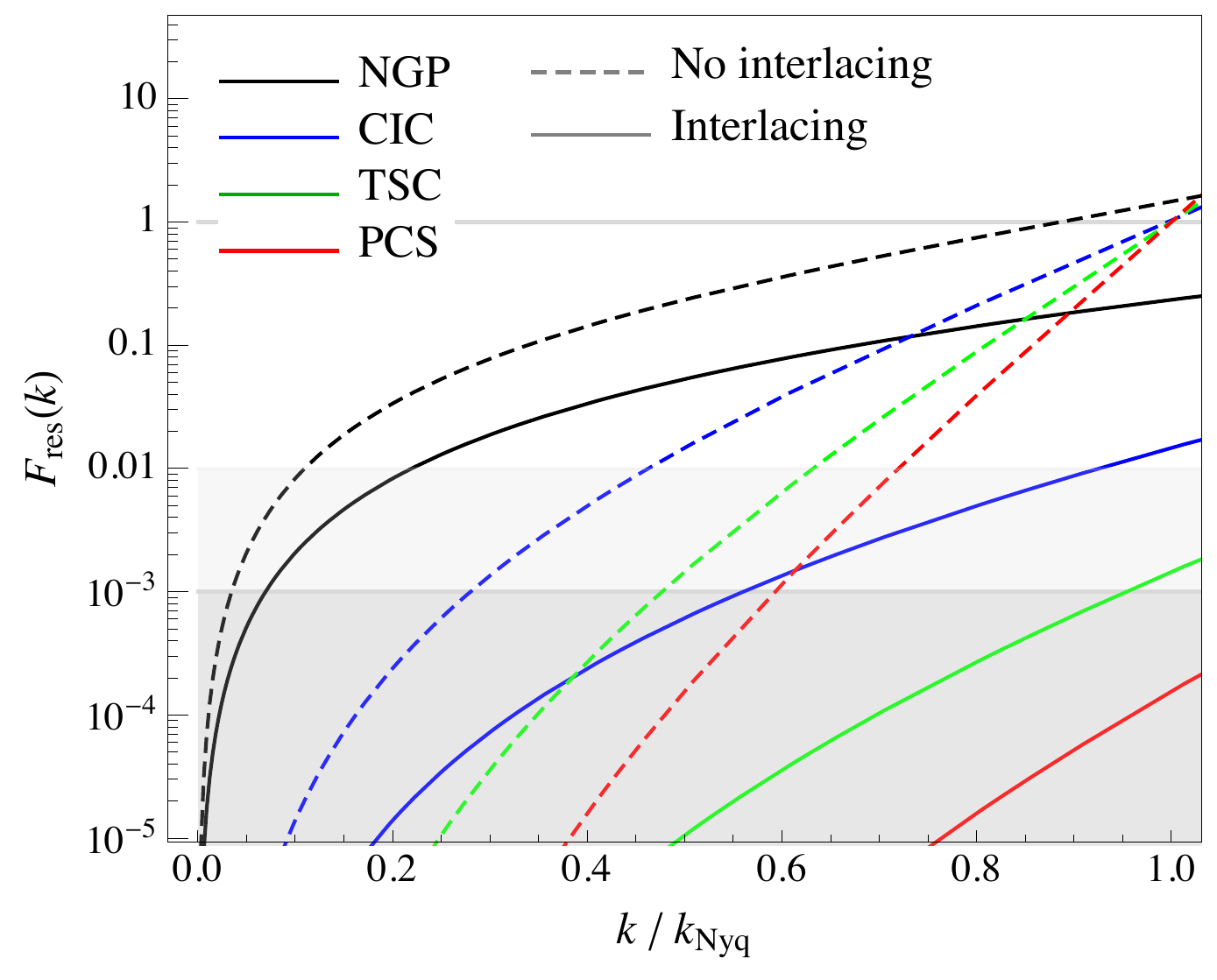}
\caption{ 
Residual factor $F_{\rm res, 1D}(k)$, in one spatial dimension, as a function of the ratio of the wavenumber $k$ to the Nyquist frequency, evaluated for the four interpolation schemes considered. Continuous and dashed curves assume respectively results with and without interlacing, \eq{eq:res1Di} and \eq{eq:res1Dnoi}, respectively.}
\label{fig:Fres}
\end{figure}

\begin{figure*}
\includegraphics[width=.8\textwidth]{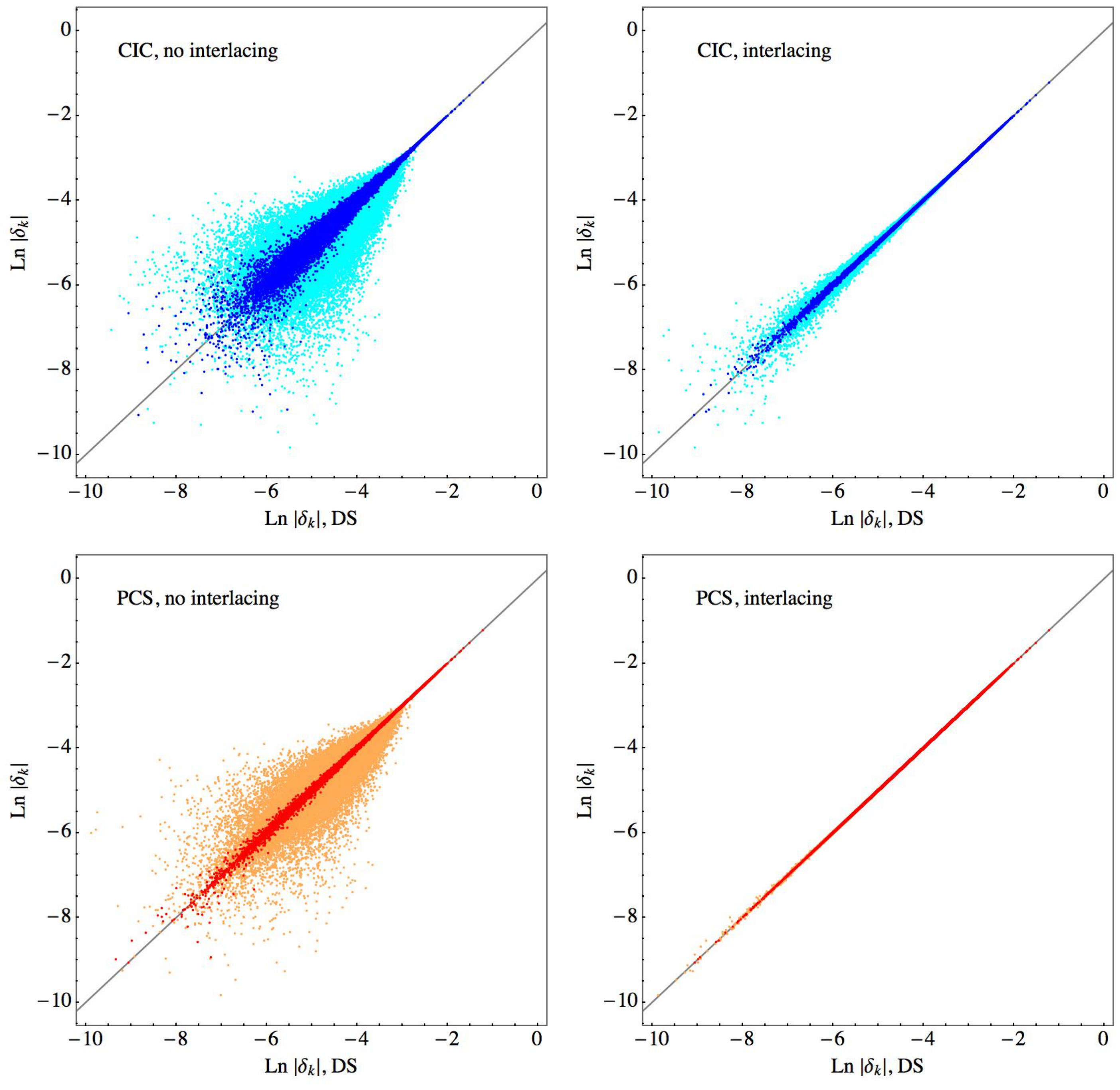}
\caption{ 
Scatter plots comparing the (log) amplitude of the Fourier-space density,  $\ln|\delta(\kv)|$ obtained by interpolation on a grid and the same quantity obtained by direct summation. Top row assumes the CIC mass assignment scheme, while bottom row the PCS mass assignment scheme. Left-hand column assumes no correction by interlacing, while the right-hand column does. In all panels light shaded (cyan and orange) points correspond to all modes in the grid of linear size $N_G=100$ while dark shaded (blue and red) points are restricted to modes of $|\kv|\le 0.7~\kN$. 
}\label{fig:amplitudes}
\end{figure*}

Fig.~\ref{fig:Fres} shows, for simplicity in one spatial dimension, the function $F_{\rm res}(k)$ for the four interpolation schemes assumed in the previous section, evaluated with and without interlacing. The figure nicely illustrates both the effects due to the choice of the interpolation scheme and due to the interlacing procedure. In the interlacing case the 1D residual function explicitly corresponds to 
\bea\label{eq:res1Di}
F_{\rm res, 1D}^{\rm ~int}(x)  &\!\!\!\! =\!\!\!\! &  \sum_{n\ne 0}\, \left(\frac{x}{x-2\,n}\right)^{2\,p}
\nn\\
& \!\!\!\!= \!\!\!\!&
 \left(\frac{x}{2}\right)^{2p}\,\left[\zeta\left(2\,p,\,x/2\right)+\zeta\left(2\,p,\,-x/2\right)\right]-2\,,
\eea
where we introduced the Hurwitz zeta function
\bea
\zeta(p,x)\;=\;\sum_{n=0}^{\infty}\frac1{(n+x)^p}\,.
\eea
Without interlacing we have instead
\bea\label{eq:res1Dnoi}
F_{\rm res, 1D}^{\rm~ no-int}(x)  \!\! \!\!  & = &\!\! \!\!  \sum_{n\ne 0}\, \left(\frac{x}{x-n}\right)^{2\,p}
\nn\\
& = &
x^{2p}\,\left[\zeta\left(2\,p,\,x\right)+\zeta\left(2\,p,\,-x\right)\right]-2\,.
\eea
We therefore obtain, in one dimension, the simple relation
\bea\label{eq:res_comp}
F_{\rm res, 1D}^{\rm ~int}(x)  \!\! \!\!  & = &\!\! \!\!  F_{\rm res, 1D}^{\rm~ no-int}(x/2)\,.
\eea
Fig.~\ref{fig:Fres} indicates in particular, as we will show with numerical tests in the next sections, that without interlacing, the aliasing contribution reaches relevant values approaching the Nyquist frequency $\kN$, regardless of the mass assignment scheme chosen. With interlacing, according to \eq{eq:res_comp}, this will happen at $k$ equal twice  $\kN$, i.e. well outside the interval of interest. 

In a similar way, we can write the bispectrum of the interlaced field as
\bea
B^G(\kv_1,\kv_2)
& \!\!\!\!=\!\!\!\! & \sum_{\nv_1,\nv_2}\,\theta_{\nv_1}\,\theta_{\nv_2}\,\theta_{-\nv_{12}}\nn\\
& & \times \,w_{\nv_1}(\kv_1)\,w_{\nv_2}(\kv_2)\,w_{-\nv_{12}}(-\kv_{12})\nn\\
& & \times \,B_{tot}(\kv_1-\nv_1\, k_s, \kv_2-\nv_2\, k_s)\,,
\eea
where the residual aliasing contributions correspond to all terms in the sums with $\nv_1\ne 0$ and $\nv_1\ne 0$.
In this case we could not find a simple expression taking advantage of the explicit form of the B-spline window functions. We will present a numerical estimate of the aliasing contribution to the bispectrum in section~\ref{sec:higherorder}.  

\begin{figure*}
\includegraphics[width=.8\textwidth]{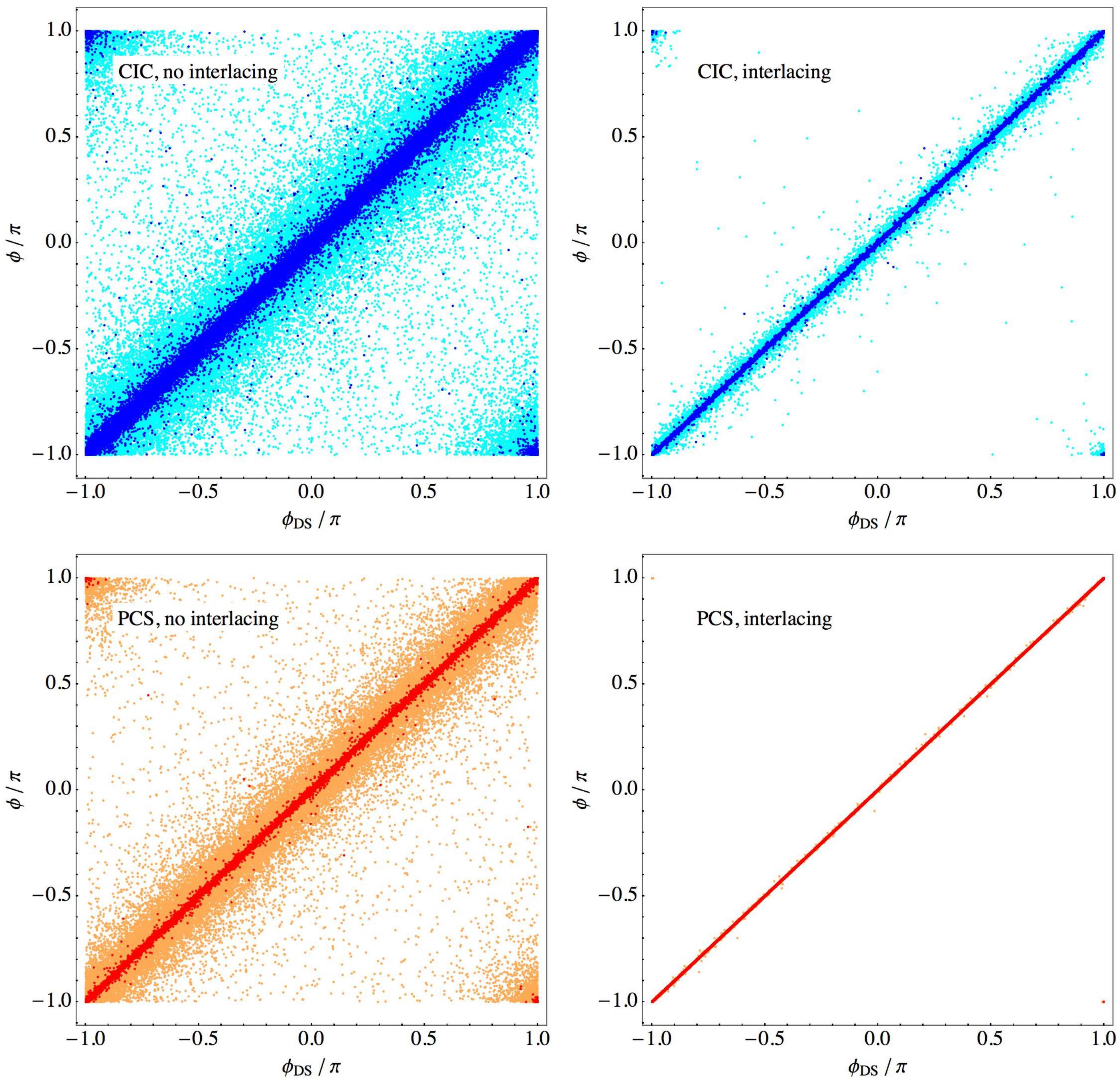}
\caption{ 
Same as figure~\ref{fig:amplitudes}, but comparing the phases of the Fourier-space mode, $\phi=\arctan\,[{\rm Im}\, \delta(\kv)\, /\, {\rm Re}\, \delta(\kv)]$.  }\label{fig:phases}
\end{figure*}

As a matter of fact, it is important to stress that the aliasing reduction is obtained, by interlacing, at the level of the density field, not after the power spectrum evaluation. As we will see, this fact ensures that the method benefits extend to all higher-order correlations that can be estimated from the FT of field itself.  Also, since interlacing only removes  the odd images, it is important to supplement this technique with a reasonable choice of interpolation kernel as discussed in the previous section to suppress the amplitude of the remaining alias images. 

\subsection{Comparison to direct summation}
\label{sec:compareDS}

The effects of aliasing are most effectively estimated comparing the density field in Fourier space obtained by a given interpolation scheme with the density field obtained by direct summation to the same grid, where they are absent. Along the lines of the tests presented in \citet{ColombiEtal2009}, in this section we consider two different particle distributions both relevant for Large-Scale Structure studies: a N-body simulation evolved to redshift zero on a relatively small box (therefore characterized by significant non-linearities) and a distribution obtained in the Zeldovich Approximation at high redshift.  Due to the significant numerical resources required by the direct summation, we will consider distributions of a relatively low number of particles.

For the $N$-body simulation, we make use of the controlled numerical experiment described in \citet{ColombiEtal2009}, run on a box of size $L=50\kMpc$ of side with $N_{\rm P}=128^3$ particles, and publicly available in the {\powmes} package as a test for the code. 

In this case, before presenting the comparison between power spectra and to underline the fact that the interlacing correction acts at the level of density field, we show a comparison between the amplitudes (as $\ln|\delta(\kv)|$, figure~\ref{fig:amplitudes}) and phases ($\phi=\arctan\,[{\rm Im}\, \delta(\kv)\, /\, {\rm Re}\, \delta(\kv)]$, figure~\ref{fig:phases}) of the density in Fourier space. Since small-scales modes dominate the scatter plot due to their larger number, we show large-scale modes characterized by  $|\kv|\le 0.7\,\kN$ by a darker shade (red or blue in the color figure), while all others are denoted by a lighter shade (orange or cyan).  While this is a rather qualitative comparison, it is evident that both the choice of the assignment scheme and the correction provided by interlacing play an important role in the reducing the scatter, and this is true for the amplitudes and phases alike. In particular, we can notice how the benefits of PCS interpolation and interlacing extend to all modes, all the way up to the Nyquist frequency, while in the other cases small scales are affected by a significant scatter.     

\begin{figure*}
\includegraphics[width=.9\textwidth]{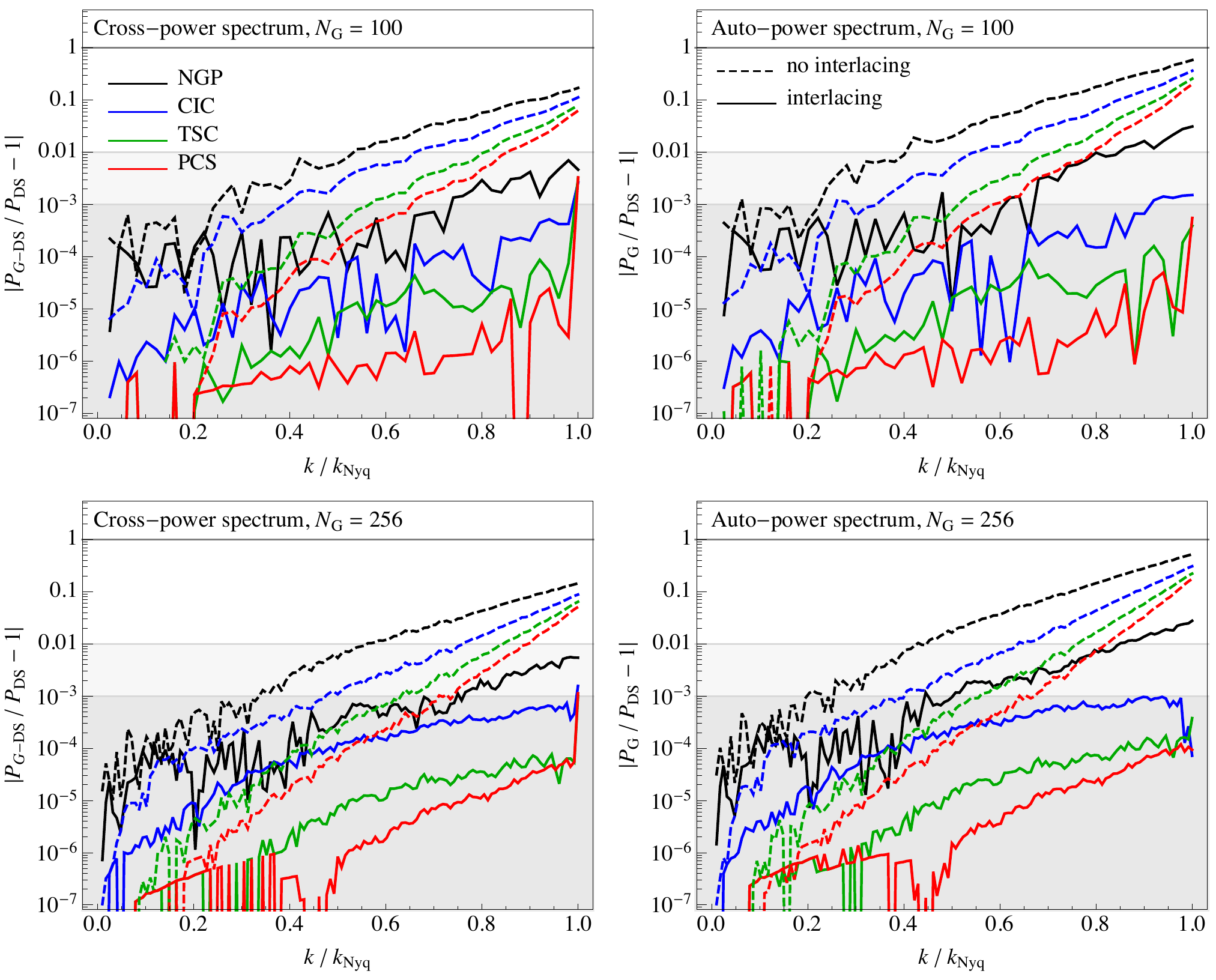}
\caption{ 
Comparison between the power spectrum measured by FFT of a grid interpolated field and the same quantity obtained by direct summation as  a function of the wavenumber in units of the Nyquist frequency. The left-hand panel show the absolute value of the relative difference between the cross-power spectrum of the grid-based, window-corrected field with the direct summation density field, $P_{\rm G-DS}\sim \langle\del^{\rm G}(\kv)\,\delta(-\kv)\rangle$ to the auto-power spectrum $P_{DS}\sim \langle|\del(\kv)|^2\rangle$. The right-hand panel shows instead the relative difference between the auto-power spectra $P_{G}\sim \langle|\del^{\rm G}(\kv)|^2\rangle$ and $P_{DS}$. Results in the top panels assume a (linear) FFT grid of $N_G=100$ while the bottom panels assume $N_G=256$. All measurements are performed on the $N$-body simulation described in \citet{ColombiEtal2009}.}
\label{fig:IL}
\end{figure*}

More quantitative results can be obtained by looking at measurements of the power spectra. Fig.~\ref{fig:IL} shows the comparison between the simulation  power spectrum obtained by different mass assignment schemes and the one obtained by direct summation. In particular, the left-hand panel shows the absolute value of the relative difference between the cross-power spectrum obtained from the grid-based (and window-corrected) density field $\del^G(\kv)$, with the one obtained by direct summation, $\del(\kv)$, that is $P_{G-DS}\sim \langle\del^G(\kv)\,\delta(-\kv)\rangle$, to the auto-power spectrum $P_{DS}\sim \langle|\delta(\kv)|^2\rangle$, in short
\bdm
\left|\frac{P_{G-DS}(k)}{P_{DS}(k)}-1\right|\,.
\edm
This quantity provides an estimate of the error induced on the density field itself by the grid assignment and aliasing.  The right-hand panel of fig.~\ref{fig:IL} shows instead the relative difference between the auto-power spectra $P_{G}\sim \langle|\del^G(\kv)|^2\rangle$ and $P_{DS}$, 
\bdm
\left|\frac{P_{G}(k)}{P_{DS}(k)}-1\right|\,,
\edm
larger roughly by a factor of two, as one can expect. Both auto- and cross- power spectra are corrected for shot-noise simply by
subtracting the inverse number density of the box. Upper panels assume an interpolation grid of linear size $N_G=100$ while lower panels
assume $N_G=256$. In all cases, results are shown as a function of the ratio of the wavenumber $k$ to the Nyquist frequency $\kN\equiv
\pi\,N_G/L$, $L$ being the size of the box. In this way it is easy to read-off the reach in the range of validity of each interpolation
scheme in terms of the relative fraction of the Nyquist. For instance, the commonly used CIC scheme, when no aliasing correction is applied,
provides a percent-level accuracy only up to 0.6~$\kN$. The immediate
consequence of this fact is that if percent accuracy is required on a
given range of scale, to avoid unwanted aliasing contributions the
interpolation grid should be $\sim 6$ ($1/0.6^3$, in 3 dimensions) times larger
than the one naively expected in the case such accuracy could be reached over all wavenumbers, resulting in much larger numerical effort. The situation is even worse if the required accuracy is at the 0.1\% level: in this case the CIC reach is $k\lesssim 0.4~\kN$. Different, higher-order interpolation schemes can marginally improve the range of scales over which a given accuracy is recovered, but they show aliasing components converging to the same value (at tens of percent level) as the Nyquist frequency is approached.  

Such behaviour is significantly reduced, instead, when the interlacing technique is applied. In this case, a 0.1\% accuracy is reached, with a CIC scheme, all the way up to the Nyquist, while PCS provides 0.01\% accuracy at the Nyquist frequency and reaches the limits imposed by single precision arithmetic ($\sim 10^{-7}$) used in this test at low wavenumbers.  

\begin{figure*}
\includegraphics[width=.9\textwidth]{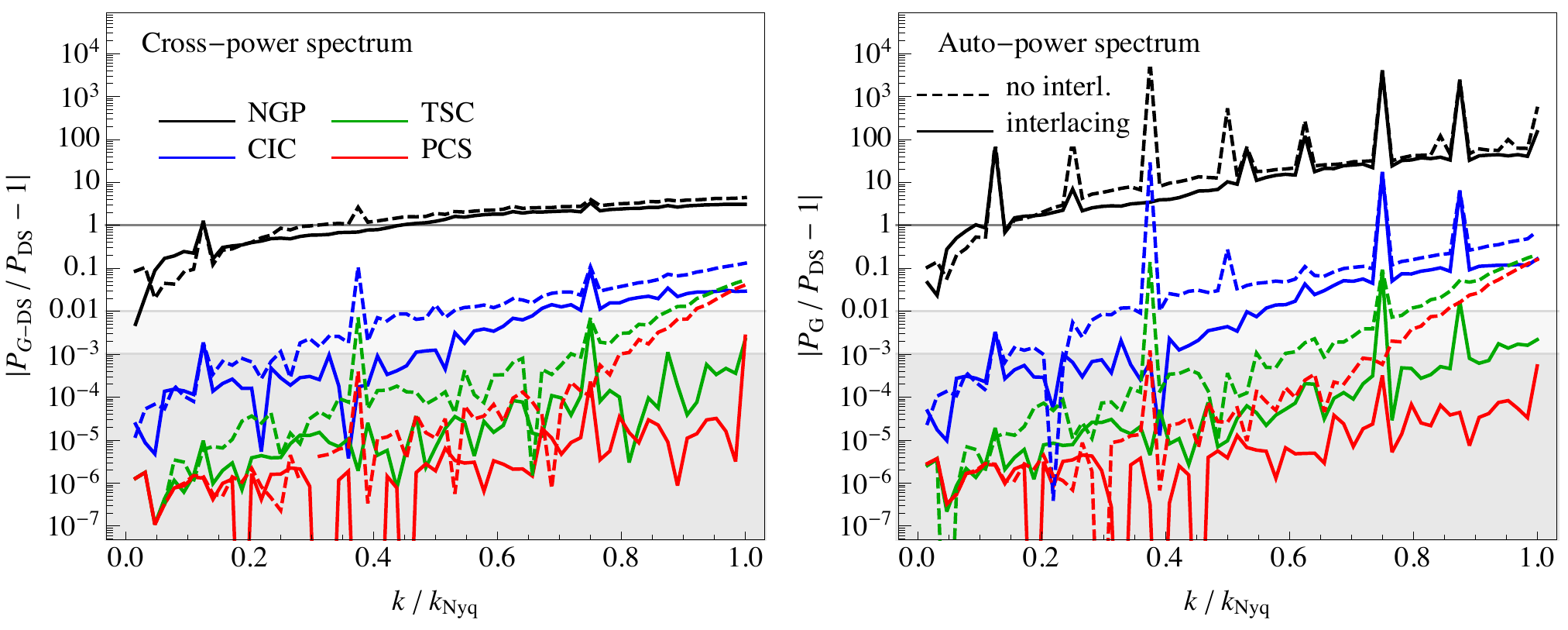}
\caption{ Same as figure~\ref{fig:IL}, but for a assuming a distribution of $360^3$ particles obtained in the Zel'dovich Approximation at redshift $z=50$, therefore similar to the initial conditions of an N-body simulation. All measurements assume a density grid of linear size $N_G=128$.}
\label{fig:ILZA}
\end{figure*}

It should be noted, however, that these results so far are specific to an $N$-body simulation at $z=0$ run on a relatively small box. Since  aliasing depends on the small-scale power it is important to consider different distributions to have a more complete picture of the problem.  As an additional, alternative test, we consider a distribution of $360^3$ particles obtained in the ZA at
redshift $z=50$ in a box of side $L=2400\kMpc$. Fig.~\ref{fig:ILZA} shows the same quantities as
Fig.~\ref{fig:IL}, but, in this case, we limit the comparison to a
grid of linear size $N_G=128$. At the large redshift assumed, with the
particle distribution corresponding to the typical  initial
conditions of an N-body simulation, the displacements from the initial
grid are very small respect to the grid scale and no shot-noise
correction is applied.  We see from this figure that NGP cannot capture the small displacements (which is not surprising) while CIC also has significant systematic errors even when interlacing (which does not help as much as in the previous distribution).   It is
therefore remarkable that, even for such a peculiar particle distribution, interlacing provides an accuracy better than $0.1\%$ level for TSC and $0.01\%$ for PCS.   

\subsection{Comparison to analytical predictions}

A further test of a power spectrum estimator would be given by a particle distribution with a known power spectrum.  Clearly, as the accuracy levels being tested are well below the percent level, the expected systematic error on such theoretical power spectrum should correspondingly be very small and this is a quite difficult goal to achieve. In principle, however, the ZA provides a particle distribution with a nonlinear power spectrum that can be computed exactly. This is given by
\be\label{eq:ZAth}
P_{ZA}(k)=\int\frac{d^3r}{(2\pi)^3}\,e^{i\kv\cdot\rv}\,\left[e^{-k^2\,\sigma_v^2 - I(\kv,\rv)} -1 \right]\,,
\ee
where $I(\kv,\rv)\equiv\int
d^3q\,(\qv\cdot\kv)^2\,\cos(\qv\cdot\rv)\,P_L(q)/q^4$ and
$\sigma_v^2\equiv I(k,0)/k^2$ \citep{BondCouchman1988,SchneiderBartelmann1995,
  TaylorHamilton1996}. In what follows, the ZA theoretical predictions
are computed assuming an infrared cutoff given by the fundamental
frequency of the box and an ultraviolet cutoff corresponding to the
Nyquist frequency of the initial particle grid. 

\begin{figure*}
\includegraphics[width=.9\textwidth]{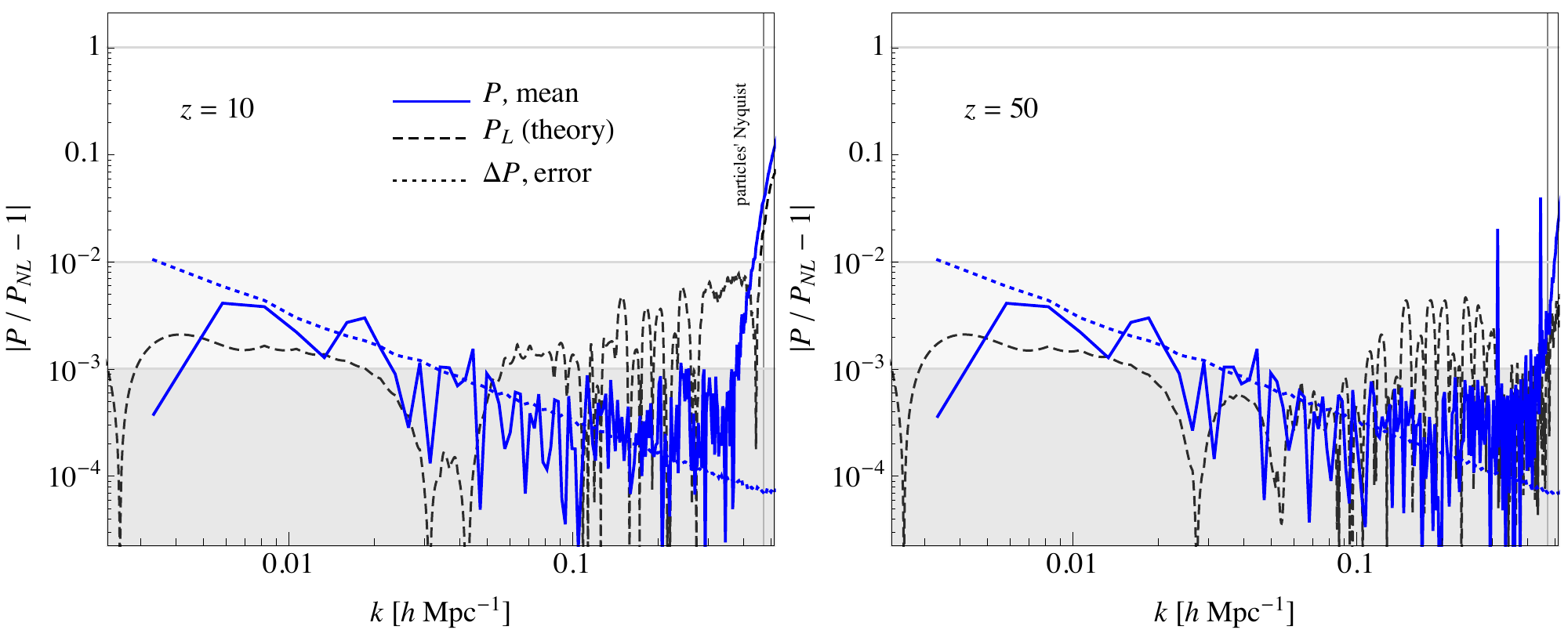}
\caption{
Absolute value of the relative difference between the mean power spectrum measured in 1000 ZA realisations (on a FFT grid of linear size $N_G=480$,  {\em blue curves}) and the nonlinear prediction of \eq{eq:ZAth}.  Also shown is the error on the mean ({\em blue, dotted curves}) and the linear prediction ({\em black, dashed curves}). Left panel assumes redshift $z=10$, right panel $z=50$. The vertical lines mark the Nyquist frequency {\em of the ZA realisations}, beyond which we expect discreteness corrections to be relevant. }
\label{fig:ZA} 
\end{figure*}

We run 1000 realisations of the ZA with $360^3$ particles in a box of side $L=2400\Mpc$, at redshifts $z=10$ and 50. Fig.~\ref{fig:ZA} shows the absolute value of the relative difference between the mean of the measured power spectra (on an FFT grid of linear size $N_G=480$, {\em blue curve}) and the nonlinear prediction. Also shown is the linear prediction ({\em black, continuous curve}) and the error on the mean ({\em blue, dotted curve}). One can notice that our measurements, done with PCS interpolation and interlacing, show the expected agreement with the nonlinear prediction while departing from the linear one despite the very small difference (of order 0.1\%)  between the two. Note that we do {\em not} subtract any shot-noise component in this case: the ZA displacements at high redshift are very small w.r.t. the initial inter-particle distance defined by the grid and no significant shell-crossing occurs at small scales. The plots are limited by the particle's Nyquist frequency, that is the Nyquist frequency of the initial ZA grid, which has a smaller value than Nyquist frequency of the FFT grid. In fact one can notice that some additional component to the measured power spectrum appears near the particle's Nyquist frequency, shown in the plots by a vertical line. This is likely due to the discrete nature of the ZA realisations and therefore arising from the interplay between the FFT grid size and the particle grid \citep[see][for a theoretical description of these effects]{Gabrielli2004, ColombiEtal2009}. We did not attempt to derive analytically the expected correction, but we compared the results obtained from realisations characterized by a different number of particles ($360^3$ and $400^3$, not shown) keeping all other variables unchanged. We did notice that the additional component near the Nyquist frequency is different for the two distributions and it is actually reduced for the one with the largest particle density.

\subsection{Comparison to {\powmes}}

The {\powmes} code \citep{ColombiEtal2009}, based on a Fourier-Taylor expansion allows for a determination of the power spectrum with arbitrary accuracy at the expenses of numerical resources. In practice, it reduces the contribution of aliasing by increasing the order $N$ of the expansion, while providing an estimate of the residuals at each order. However, it thus requires a number of FFTs at each order given by $N_{FFT}=(N+3)! / (N!\,3!)$, so that the total computational cost is ${\mathcal O}(N_{FFT}\times N_G^3\,\ln\,N_G)$, where $N_G$ is the linear size of the grid. 

\begin{figure*}
\includegraphics[width=.9\textwidth]{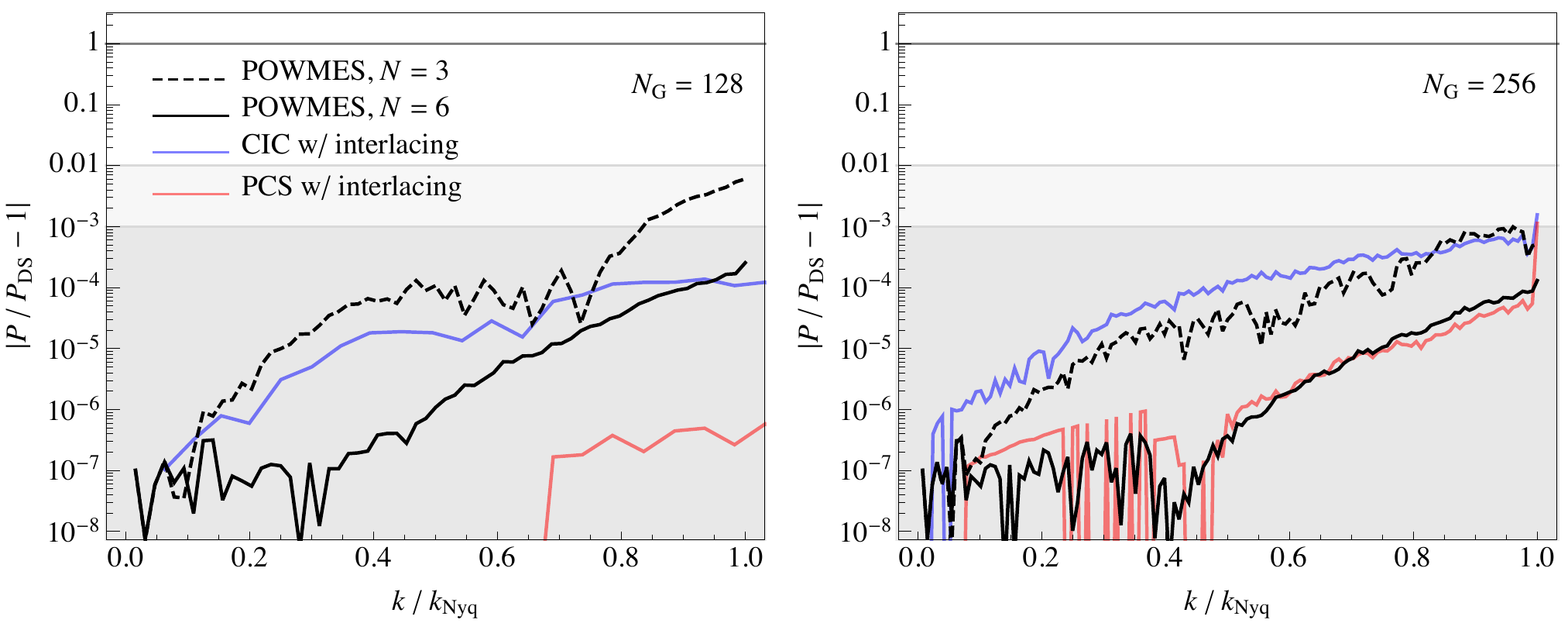}
\caption{
Relative difference between the {\powmes} ({\em black curves}) measurements and our measurements ({\em blue and red curves}) with the direct summation result performed over the N-body simulation described in section~\ref{sec:compareDS}. Dashed and continuous black curves assume, respectively, the order $N=3$ and $6$ in the Taylor expansion of {\powmes}. Blue and red continuous curves correspond respectively to the CIC and PCS interpolation schemes, both with interlacing. The left panel assumes a grid of size $N_G=128$, while the right panel assumes
$N_G=256$. }
\label{fig:powmes} 
\end{figure*}

Figure~\ref{fig:powmes} shows a comparison between measurements of the {\powmes} code and CIC and PCS interpolation with the
interlacing scheme proposed in this paper, in terms of their relative difference with the direct summation measurement of the power spectrum. For the particle distribution we assume again the test N-body simulation of \citet{ColombiEtal2009}. 

Within the {\powmes} results themselves one can notice a clear improvement when comparing different orders in the Fourier-Taylor
expansion (for a given grid size). This comes, however at a large computational cost. In this example, the expansion orders $N=3$ and $N=6$ for {\powmes} correspond to a number of FFTs of $N_{FFT}=20$ and 84. Remarkably a comparable improvement (or better, depending on $N_G$) can be achieved moving from CIC to PCS interpolation while performing interlacing. In this case, we only require two FFTs.

\section{Higher-order correlations}
\label{sec:higherorder}

A positive aspect of correcting for aliasing at the level of the density field itself consists in extending its benefits to the estimation of all higher-order correlation functions, not just to power spectrum measurements. In this section we consider the case of the bispectrum, i.e. the 3-point function in Fourier space.   

A general expression for the estimator of the bispectrum function of a triplet of wavenumbers $\kv_1$, $\kv_2$ and $\kv_3$ in the case of a simple geometry is given by
\bea\label{eq:Best}
\hat{B}(k_1,k_2,k_3) & \equiv & \frac{1}{V_B}\int_{k_1}\!\!\de^3q_1\int_{k_2}\!\!\de^3 q_2\int_{k_3}\!\!\de^3 q_3\,\delta_D(\qv_{123})\nn\\
& & \times\delta_{\qv_1}\,\delta_{\qv_2}\,\delta_{\qv_3}\,,
\eea
where the integrals are assumed to be over a spherical shell of size $\Delta k$, with radius centred at $\qv_i=\kv_i$, that is
\be
\int_{k}\!\!\de^3q\equiv\int_{k-\Delta/2}^{k+\Delta k/2}\de q\,q^2\int\de\Omega\,,
\ee 
and where the normalization factor $V_B$, given by
\be\label{eq:VBest}
V_B(k_1,k_2,k_3)\equiv\int_{k_1}\!\!\de^3q_1\int_{k_2}\!\!\de^3 q_2\int_{k_3}\!\!\de^3 q_3\,\delta_D(\qv_{123})\,,
\ee
 is proportional to the number of fundamental triplets $\qv_1$, $\qv_2$ and $\qv_3$ that can be found in the triangle bin defined by the wavenumbers $k_1$, $k_2$ and $k_3$, with width $\Delta k$. In both the integrals of equation \eq{eq:Best} and \eq{eq:VBest}, the Dirac delta ensures that the triplet of $\qv_i$ actually forms a closed triangle. 
 
\subsection{A fast bispectrum estimator}

The implementation of the expression in \eq{eq:Best} can take advantage of the integral representation of the Dirac delta, so that \citep[see][for extensions to  redshift-space]{Scoccimarro2015}
\bea\label{eq:Best2}
\hat{B}(k_1,k_2,k_3) & =  & \frac{1}{V_B}\int \frac{\de^3x}{(2\pi)^3}\int_{k_1}\!\!\de^3q_1\int_{k_2}\!\!\de^3 q_2\int_{k_3}\!\!\de^3 q_3\nn\\
& &\times \delta_{\qv_1}\,\delta_{\qv_2}\,\delta_{\qv_3}\,e^{i\,\qv_{123}\cdot\xv}\nonumber \\
& = & \frac{1}{V_B}\int \frac{\de^3x}{(2\pi)^3}\,\prod_{i=1}^3\,I_{k_i}(\xv)\,,
\eea
where the integrals 
\be\label{eq:IntI}
I_{k}(\xv)\equiv \int_{k}\!\!\de^3q\,\delta_{\qv}\,e^{i\,\qv\cdot\xv}\,,
\ee
whose number is equal to the number $N_k$ of $k$-bins we are interested in, can be evaluated as FFTs (scaling as $N_G^3\,\ln\,N_G^3$) and stored in memory. The final integral in \eq{eq:Best2} corresponds to an $\Oc(N_G^3)$ process. The normalisation factor $V_B(k_1,k_2,k_3)$ can also be evaluated in a similar way. 

Notice that because of the exponential $e^{i\qv_{123}\cdot\xv}$ in \eq{eq:Best2}, this estimator cannot be applied to all values of $k_i$ up to the Nyquist frequency of the box. In fact, for a translation of the wavenumbers $\qv$ given (in one dimension) by $q_i \rightarrow q_i + \frac{2\pi}{L}\,\frac{N_G}{3}$ where $N_G$ is the grid size and $L$ the linear size of the box, we have $q_{123} \rightarrow q_{123} + \frac{2\pi\,N_G}{L}$ and with $x=m\,L/N_G$ the exponential factor would be invariant. It follows that the largest value for the wavenumbers $k$ turns out to be $k_{\rm max}=k_f\,N_G/3$, $k_f$ being the fundamental frequency rather than $k_{\rm max}=k_f\,N_G/2$, the Nyquist, as it is the case for the power spectrum estimator.
 
This implementation of this estimator has been considered already in the past in the context of the Large-Scale Structure \citep{Scoccimarro2000B, FeldmanEtal2001, ScoccimarroEtal2001B}. Note that the expression in \eq{eq:Best2} is analogous to the estimator of the microwave anisotropy bispectrum considered by \citet{KomatsuEtal2002}.

\subsection{Aliasing contributions on the bispectrum}

Despite the efficient  bispectrum estimator described above, it is still crucial to limit as much as possible numerical requirements starting with the size of the FFT grid. This is true, in particular, regarding the memory requirement since the implementation of \eq{eq:Best2}  benefits from storing in memory the integrals of \eq{eq:IntI}, i.e. a number of arrays of size $N_G^3$ equal to the number of $k$-bins requested.  The interlacing scheme described in the previous section therefore results is particularly useful since it allows to efficiently use the  $N_G^3$ FFT grid, with the aliasing contribution under control all the way up to the Nyquist frequency.    

\begin{figure*}
\includegraphics[width=.9\textwidth]{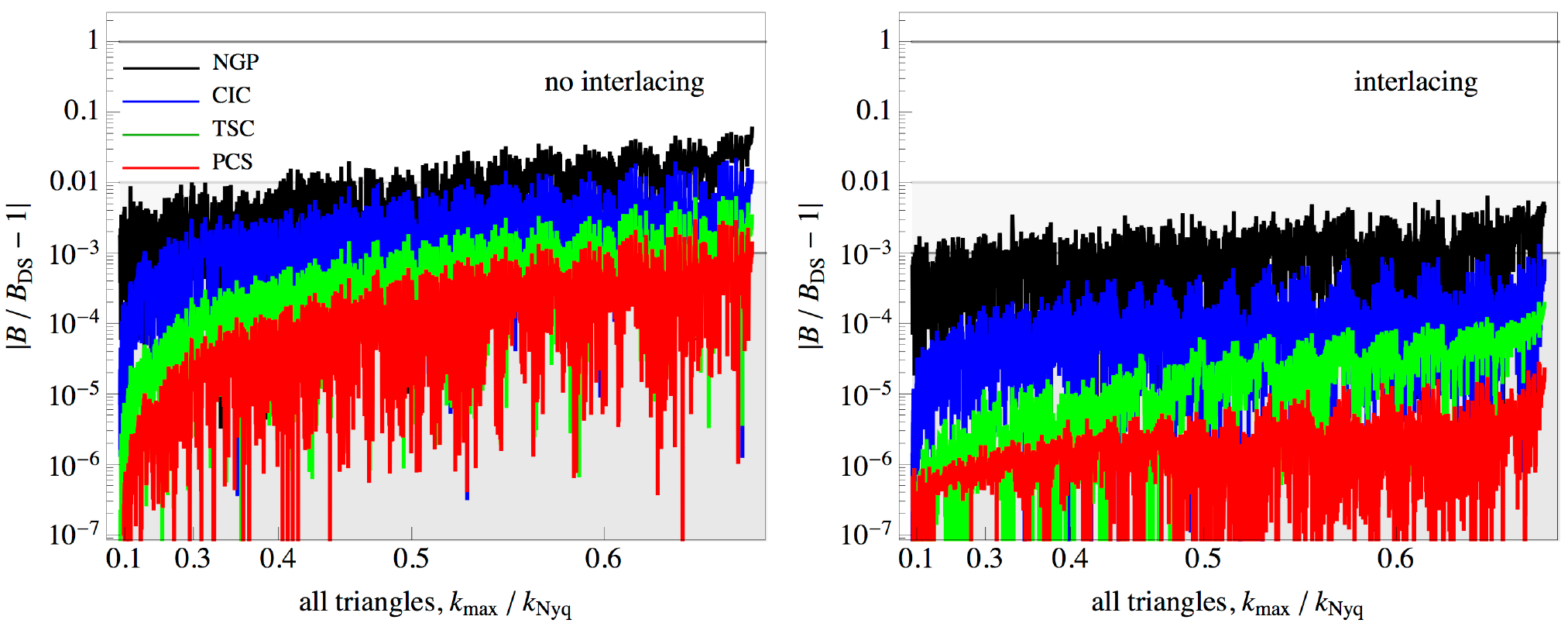}
\caption{ 
Aliasing effects on the bispectrum. Both panels show the relative difference between the bispectrum measured with distinct interpolation schemes ({\em NGP, CIC, TSC, and PCS, as black, blue, green and red curves, or different shades top to bottom}) with respect to the bispectrum measured from the density field estimated by direct summation. Results on the left-hand panel do not implement interlacing, while the results on the right-hand panel do. We show all triangular configurations with triplets $(k_1,k_2,k_3)$ measured in units of the fundamental frequency, $k_f$, up to $2\kN/3$ (see text for explanation). On the $x$-axis are shown the values of $k_{max}\equiv \max(k_1,k_2,k_3)$ in units of the Nyquist frequency. }
\label{fig:ILbisp}
\end{figure*}

Figure~\ref{fig:ILbisp} shows the relative difference between the measurements of the matter bispectrum obtained from the density interpolated on a grid to the measurements of the same quantity obtained from the density evaluated by direct summation. We assume here the $128^3$ particle $N$-body simulation already considered for the power spectrum tests. All measured triangular configurations are shown, ordered by increasing values of $k_1\ge k_2\ge k_3$, $k_1$ therefore representing the largest value of the triplet. For the reason discussed above, the largest value accessible by the estimator is $k=2\,\kN/3$. The left-hand panel compares different interpolation schemes without aliasing corrections, while the right-hand panel shows the same results including interlacing.  In the latter case, already for the CIC  scheme, the aliasing contribution is kept below the $0.1\%$ level, while $0.001\%$ is achieved with PCS interpolation. If no aliasing correction is implemented the error can reach the few percent level already at $k\sim 0.5 \,\kN$. 

Not surprisingly, these results are largely consistent with those obtained for the power spectrum. 

\section{Conclusions}
\label{sec:conclusions}

Standard FFT-based, power spectrum estimators routinely employed in the analysis of redshift surveys data sets as well as $N$-body simulations, suffer from spurious contributions due to aliasing. This well-known problem is often circumvented by limiting the range of scales (wavenumbers) over which the desired accuracy could be achieved. The drawback, clearly, is the inefficient use of numerical resources as this approach clearly requires larger FFT grids than naively expected. This  becomes even more problematic for higher-order correlations.

Aliasing takes the form of a sum over images, each of them corresponding to the  Fourier content displaced by increasing multiples of the sampling frequency of the grid. The strength of the aliasing contributions depends on both the interpolation kernel that goes from objects to density perturbations on the grid and the small-scale modes not supported by the grid. We quantified  the systematic error on the power spectrum estimation due to aliasing in two particle distributions relevant for Large-Scale Structure studies: an N-body simulation describing the (nonlinear) matter field at $z=0$ and the distribution obtained by particles displaced from a regular grid according to the Zel'dovich Approximation at $z=10,50$. 

By comparing our results to the (aliasing-free) estimation by direct summation, we showed that in the $N$-body example the resulting error on the power spectrum estimation can be larger than 1\% already at scales roughly above half the Nyquist frequency of the grid  when using the standard Cloud-In-Cell (CIC) interpolation. If the required accuracy is below the $0.1\%$ level, a reasonable requirement for cosmological applications aimed at the detection of small effects possibly due to non-standard physics, the usable range of scales becomes as restrictive as $k\lesssim 0.4\kN$.  The problem is even more severe when a distribution like the one obtained by ZA at large redshift is considered: in this case an accuracy of 1\% (0.1\%) is only achieved for wavenumbers $k\lesssim 0.3~ (0.2)~\kN$. 

We then revisited a method to reduce the aliasing contribution based on an interlacing technique \citep[see, e.g.][]{HockneyEastwood1981,Couchman1991}. We showed that using two interlaced grids removes odd images, which include the dominant contribution to aliasing.  We demonstrated that this technique is able to reduce the aliasing contribution  below the 0.1\% level for practically all wavenumbers, all the way to the Nyquist frequency of the grid, when combined with interpolations schemes of order higher than the CIC. This is true for both the examples mentioned above as well as for a comparison of the mean power spectrum measured from 1000 ZA realisations with the exact theoretical prediction. For the interpolation schemes we considered the third order Triangular Shaped Cloud (TSC) scheme and the  fourth order Piecewise Cubic Spline (PCS), which are increasingly smooth schemes constructed by convolving a top-hat function with itself an increasing number of times. 

While higher-order schemes have the disadvantage of an increased time required to perform the particle assignment (without affecting the necessary memory), the payout in terms of systematic errors in Fourier coefficients can be substantial. At a fixed required systematic error, the algorithm proposed here can actually be more efficient than the standard CIC method. To make systematics in the power spectrum negligible, e.g. $0.01\%$, one requires from CIC a grid size five times larger in each dimension than PCS with interlacing, which results in eight times more computational cost (and sixty-two times more memory). Similarly due to the increase in grid size, the FFT step is correspondingly slower in the CIC case. 

We remark that the aliasing correction implemented via interlacing is performed on the density field itself and have shown that combining PCS with interlacing gives very accurate Fourier amplitudes and phases of density perturbations compared to the standard CIC method. It therefore improves the estimation of the power spectrum {\em as well as} any higher-order correlation functions evaluated from such density.   We showed that this is indeed the case for the bispectrum, by performing comparison to the direct summation results, as done for the power spectrum. Other approaches to limit aliasing effects in the estimation of the power spectrum as the one proposed by \citet{Jing2005} cannot be easily extended to higher-order correlation functions.

We believe that the method presented here represent a numerically efficient way to reduce aliasing effects in the measurement of  correlation functions of particle distributions in Fourier space,  relevant in a variety of cosmological investigations. A code implementing the algorithms here described is available at \texttt{https://github.com/sefusatti/PowerI4}.

\section*{Acknowledgments}

We thank the referee, St\'ephane Colombi, for making the {\powmes} code public and for useful comments and suggestions that improved the presentation. ES thanks Julien Bel for many discussions and the NYU Dept. of Physics and the Institut des Ci\`encies de l'Espai for hospitality during the completion of this project. MC has been partially funded by AYA2013-44327 and acknowledges support from the Ramon y Cajal MICINN program. RS was partially supported by NSF grant AST-1109432, and HMPC was supported by NSERC.

\bibliography{Bibliography}

\end{document}